\documentstyle[twocolumn,aps,floats,babel,spanish,prl,latexsym,amsmath,amssymb,eqsecnum]{revtex}

\begin{document}

\draft \tolerance = 1000

\setcounter{topnumber}{1}
\renewcommand{\topfraction}{1}
\renewcommand{\textfraction}{0}
\renewcommand{\floatpagefraction}{1}
\newcommand{\br}{{\bf r}}

%Fixing abstract in twocolumn mode
\twocolumn[\hsize\textwidth\columnwidth\hsize\csname@twocolumnfalse\endcsname

\title{{\rule{17.5cm}{.15mm}{\bfseries{\\\vspace{3mm}GRAVITY  QUANTA,  ENTROPY  AND  BLACK  HOLES\\\rule{17.5cm}{.15mm}}}}}
\author{Antonio Alfonso-Faus}
\address{E.U.I.T. Aeron\'autica\\
Plaza Cardenal Cisneros s/n  ~ Madrid 28040 ~ SPAIN\\
E-mail:aalfonso@euita.upm.es}
\maketitle
\vspace*{0.2cm}
\centerline{\em Accepted for publication in ``Physics Essays". Vol. \textbf{12} Nº4 December 2000}
\vspace*{0.cm}
\begin{abstract}
We propose the use of a gravitational uncertainty principle for gravitation.
We define the corresponding gravitational Planck's constant and the
gravitational quantum of mass. We define entropy in terms of the quantum of
gravity with the property of having an extensive quality. The equivalent 2nd
law of thermodynamics is derived, the entropy increasing linearly with
cosmological time. These concepts are applied to the case of black holes,
finding their entropy and discussing their radiation.\medskip

Nous proposons ici l'utilisation d'un principe d'indet\'ermination
gravitational pour la gravitation. Nous d\'efinissons ainsi la
correspondante constante gravitationale de Planck et le quant gravitational
de masse. Nous d\'efinissons l'entropie en fonction des quants de gravit\'e
avec la propi\'et\'e de qu'elle ait une qualit\'e extensive. L'equivalente
deuxi\'eme loi du thermodynamique elle se derive ici, l'entropie cro\^itant
lin\'eairement avec le temps cosmologique. Ceus-ici concepts nous les
appliquons ou cas des trous noires trouvant son entropie et discutant son
radiation.\medskip

\end{abstract}
\vspace{.4cm}

%Fixing abstract in twocolumn mode
]
\section{{\bf INTRODUCTION}.}

The Heisenberg uncertainty principle is the basis for quantum mechanics,
through the use of Planck's constant. On the other hand, quantum gravity is
not yet an established theory. One would like to have a parallel treatment
for gravity with a gravitational Planck's constant for this case. We derive
here the value for such a constant. It turns out to be of the same order of
magnitude as the one proposed by Zecca(\cite{Z}) but following a very
different argument. A step forward in quantum gravity would have important
cosmological implications, and here we link black hole properties with the
size of the Universe, an important cosmological parameter. The hope is to
get some light in the direction of a quantum treatment of gravitation.%
\bigskip\

Based upon the gravitational uncertainty principle we derive the value for
the quantum of gravity. Using this we define the entropy of any system, from
a gravitational point of view, and get an equivalent to the 2$^{nd}$ law of
thermodynamics. Entropy is defined in a cosmological context. In particular,
our definition for entropy keeps its extensive quality. One can then derive
the primordial black hole entropy. In fact, for the entropy of any black
hole we get a result in contrast with the Bekenstein (\cite{B}) and Hawking (%
\cite{H}) formulation. In our case the extensive property for the entropy is
kept valid.\bigskip\

Finally we prove that with these new concepts the radiation from black holes
is many orders of magnitude less than the Hawking radiation formula.

\section{\bf \ A GRAVITATIONAL UNCERTAINTY PRINCIPLE.}

Elementary particles of typical mass $m_p$ have a characteristic size $r_p$
of the order of the Compton wavelength
\begin{equation}
\label{eq1}r_p\approx \frac \hbar {m_pc}\approx 2\cdot 10^{-13}cm
\end{equation}
where $\hbar $ is Planck's constant and $c$ the speed of light. For the
numerical value we have used the proton mass. If these particles emit
gravity quanta of characteristic wavelength $\lambda $ one must have $%
\lambda \approx r_p$ . i.e., one must have the same order of magnitude
between the antenna size and the wavelength of the emitted radiation. Then,
the gravity quantum equivalent mass $m_1$ must be related to a gravitational
Planck's constant $H$ in the following way,
\begin{equation}
\label{eq2}\lambda \approx r_p\approx \frac \hbar {m_pc}\approx \frac
H{m_1c}
\end{equation}
where $\hbar \gg H$ and $m_p\gg m_1.$ For gravity quanta one expects for $H$
to play the role of $\hbar $. Let us see the rate of emission of gravity
quanta. The characteristic speed must be the speed of light, and sizes being
of the order of $r_p$ as in (\ref{eq1}) the rate of emission $R=1/\tau $
must be
\begin{equation}
\label{eq3}R=\frac 1\tau =\frac c{r_p}\approx \frac{m_pc^2}\hbar \approx
\frac{m_1c^2}H
\end{equation}
Let us introduce gravitation explicitly in the picture. We know that the
gravitational field has negative energy. In fact the gravitational potential
between two masses $M$ and $m$ at a distance $r$ is
\begin{equation}
\label{eq4}-\frac{GMm}r
\end{equation}
When the distance $r$ tends to infinity this energy tends to zero as it
should. As the distance $r$ becomes less and less, the gravitational
potential decreases, becoming more negative. Applying an order of magnitude
estimate for the energy equation, considering a hydrogen atom of total
energy $E$ and rest energy $E_0$ falling radial in the gravitational field
of a much larger mass $M$ one has
\begin{equation}
\label{eq5}E-\frac{GM\left( \frac E{c^2}\right) }r=E_0
\end{equation}
As the atom falls in the gravitational field of $M$, i.e. approaches $M$
more and more if geometrically possible ($M$ concentrated enough), $E$
becomes relativistic due to the increase in kinetic energy and one can then
neglect the term $E_0$ in (\ref{eq5}). Hence, there is a limit $r_{bh}$ in
the distance of approach given by
\begin{equation}
\label{eq6}r_{bh}\approx \frac{GM}{c^2}
\end{equation}
This is the order of magnitude of the gravitational radius of the mass $M$.
It is the limit of approach that any body can have going towards the mass $M$%
. It is well known that if $M$ has a size of this order or less, it is then
a black hole. From an outside observer's point of view the body never
reaches $M$. It would take an infinite time for that. From the point of view
of the body time runs differently and it crosses the gravitational radius in
a never returning way. We see then the meaning of the gravitational radius
as a characteristic size of the corresponding black hole when the mass is
all inside this radius.\bigskip\

Emission of gravity quanta implies emission of negative energy, and
therefore an increase in the positive mass of the source, the particle.
Having found the rate of emission (\ref{eq3}) of gravity quanta of mass $m_1$
one can equate the rate of mass emission to the ratio $m_p/t$, where $t$ is
the age of the Universe, i.e.,
\begin{equation}
\label{eq7}\frac{m_1}\tau \approx \frac{m_p}t
\end{equation}
Using (\ref{eq3}) and (\ref{eq7}) one gets the equivalent set%
$$
\frac{m_1^2c^2}H\approx \frac{m_p}t
$$
\begin{equation}
\label{eq8}m_1\approx \frac H\hbar m_p
\end{equation}
and solving for $m_1$ and $H$ we finally have%
$$
m_1\approx \frac \hbar {c^2t}=\frac \hbar {m_pc}\frac 1{ct}m_p\approx
10^{-41}m_p
$$
\begin{equation}
\label{eq9}H\approx \frac \hbar {m_pc}\frac 1{ct}\hbar \approx 10^{-41}\hbar
\end{equation}
where we have used the present age of the Universe $t\approx 1.5\cdot
10^{10} $ years.\bigskip\

In 1975 Zecca(\cite{Z}) proposed a gravitational Planck's constant, derived
from a very different argument, arriving at a value of this order of
magnitude for $H$ too. The quantum world is governed by Planck's constant.
But quantum mechanics does not include gravitation. A quantum gravity
approach should have the gravitational Planck's constant derived here as the
fundamental constant for gravitation, particularly for quantum radiation.
The parallel gravitational uncertainty principle must be formulated with the
corresponding gravitational Planck's constant $H$. For example, the
radiation of a quantum of gravity of energy $E_0=m_1c^2$ must be governed by
the gravitational uncertainty relation%
$$
\tau E_0\approx H
$$
i.e.
\begin{equation}
\label{eq10}\frac \hbar {m_pc^2}m_1c^2\approx H
\end{equation}
which is equivalent to the relation found in (\ref{eq9}).

\section{\bf ENTROPY}

We may see now that entropy and mass are proportional if we take the
emission of $m_1$, the quantum of gravity, as the production of a unit of
entropy. There is a deep physical meaning here. We are claiming that the
basic quantum of gravitation is $m_1$, and that it represents an entity by
itself: we just count the number of these emitted entities to get the
entropy. Hence we are counting items of deep physical significance, the
gravity quanta, to get the entropy. Since $m_p$ is of the order of $%
10^{41}m_1$ we can take the number of emitted gravity quanta as the entropy
of the particle, about $10^{41}$. This is the number of parts that the
particle has contributed to the Universe in the form of gravity quanta. This
means that the entropy of the seeable Universe (size $ct$) is this number
times the number of particles in it (we approximately know the density and
size of the Universe, so that we know the equivalent number of protons): $%
10^{41}\cdot 10^{79}=10^{120}$ . Entropy is an extensive quantity, and under
this view it is just the total mass of the system in $m_1$ units: it is not
a question of expressing entropy in a particular unit. It is a question of
taking the emission of gravity quanta as the accounting for the creation of
entropy. Of course we are expressing the entropy $S$ in non-dimensional
form, $S/k$, where $k$ is the Boltzmann constant. We can express the entropy
of the Universe as the product of he number of parts each particle has
contributed to the Universe times he number of particles $N_p$:
\begin{equation}
\label{eq11}\frac{S_u}k\approx t\frac{m_pc^2}\hbar N_p
\end{equation}
In this sense the entropy of the Universe is the number of ``tics'' of each
particle times the number of particles in the Universe. Entropy increases
linearly with time and is very high today just because the Universe is old.
In standard cosmology there is a problem with the entropy. Usually entropy
is taken as the number of photons in the background blackbody radiation (at $%
2.73^{\circ }K$) about $10^{88}$; no one knows where it comes from or why is
so high. Now we know where the entropy comes from. Here we see that the
entropy of the Universe is a linear function of time. That is why is so high
today. The argument for the entropy of a system to be the maximum number of
parts it consists of is here reproduced: the maximum number of parts in the
Universe is the number of gravity quanta. Hence, any mass $M$ has an entropy
$S$ as
\begin{equation}
\label{eq12}\frac Sk\approx \frac M{m_1}=M\frac{c^2t}\hbar =\frac M{m_p}
\frac{ct}{r_p}
\end{equation}
i.e. the number of particles of mass $m_p$ times the factor $ct/r_p$ $%
\approx 10^{41}$ today.

\section{\bf PRIMORDIAL BLACK HOLE ENTROPY.}

We may define a primordial black hole of mass $M_p$ as the black hole that
has the size of an elementary particle, $r_p$:
\begin{equation}
\label{eq13}\frac{GM_p}{c^2}=r_p=\frac \hbar {m_pc}
\end{equation}
Hence the mass of a primordial black hole so defined is
\begin{equation}
\label{eq14}M_p=\frac{\hbar c}{Gm_p}=\frac{\hbar c}{Gm_p^2}m_p\approx
10^{38}m_p\approx 10^{79}m_1
\end{equation}
Should these objects be responsible for the dark matter in the Universe
there would be about $10^{41}$ of them. The entropy of one of them is
clearly about $10^{79}$, the number of gravity quanta emitted.\bigskip\

The entropy of an elementary particle $S_p$, a primordial black hole $%
S_{pbh} $ and the Universe $S_u$ are then given by%
$$
\frac{S_p}k\approx 10^{41}
$$
\begin{equation}
\label{eq15}\frac{S_{pbh}}k\approx 10^{79}
\end{equation}
$$
\frac{S_u}k\approx 10^{120}
$$

\section{\bf THE BLACK HOLE ENTROPY.}

Black hole entropy as related to mass was theoretically established many
years ago (Bekenstein\cite{B}, Hawking\cite{H}). Its value was given in
terms of the area of the event horizon, the square of the gravitational
radius of the black hole of mass $M_{bh}$, in units of Planck's length. We
know that from the physical parameters speed of light $c$, Planck's constant
$\hbar $ , and the gravitational constant $G$, one can define Planck's units
as%
$$
l_{*}=\left( \frac{G\hbar }{c^3}\right) ^{\frac 12}=1.6\cdot 10^{-33}cm
$$
\begin{equation}
\label{eq16}m_{*}=\left( \frac{hc}G\right) ^{\frac 12}=2.2\cdot 10^{-5}gr
\end{equation}
$$
t_{*}=\left( \frac{G\hbar }{c^5}\right) ^{\frac 12}=5.4\cdot 10^{-44}s
$$

Then, the Bekenstein-Hawking formula for the entropy of a black hole can be
expressed as
\begin{equation}
\label{eq17}\frac{S_{bh}}k\approx \left( \frac{GM_{bh}}{c^2}\right) \cdot
\frac 1{l_{*}^2}\approx \frac G{\hbar c}M_{bh}^2
\end{equation}
However, there are objections to this expression as given by Dunning-Davies
and Lavenda(\cite{D}). In particular it is argued that this expression is
not extensive. We see here that the Bekenstein-Hawking expression (\ref{eq17}%
) is different from the one we propose in (\ref{eq13}). If we apply (\ref
{eq17}) to the case of the primordial black hole we get an entropy of about $%
10^{38}$. In our case this number is $10^{79}$. The two concepts are very
different indeed. Our expression (\ref{eq12}) is extensive, as it should be
for entropy.\bigskip\

One of the main ideas of the Bekenstein-Hawking findings on the entropy
associated to a black hole is the horizon area. This means that entropy can
be linked to the area of the black hole (the horizon area) as measured in
Planks units. A Planck cell is of fundamental meaning in the vacuum concept.
Again, we are taking the entropy in dimensionless units, $S/k$. We can count
the number of gravity quanta or the number of Plancks cells in the horizon
area. In this way we can find an equivalent area $A_e$ in Plancks units
(Bekenstein-Hawking idea in expression (\ref{eq17})) to our relation in (\ref
{eq12}). Then one has now the new area $A_e$ that in Plancks units expresses
the value of the entropy:
\begin{equation}
\label{eq18}\frac{A_e}{l_{*}^2}=\frac{M_{bh}}{m_p}\frac{ct}{r_p}\approx
M_{bh}\cdot \frac{c^2t}\hbar
\end{equation}
and using (\ref{eq16}) we get for the equivalent area
\begin{equation}
\label{eq19}A_e\approx \frac{GM_{bh}}{c^2}ct
\end{equation}

The equivalent area has a size equal to the geometric mean of the
gravitational radius of the black hole, $GM_{bh}/c^2$, and the size of the
Universe ct. Hence, our entropy expression can be defined in this way as
\begin{equation}
\label{eq20}\frac{S_{bh}}k\approx \frac{A_e}{l_{*}^2}\approx \frac{\frac{%
GM_{bh}}{c^2}\cdot ct}{l_{*}^2}
\end{equation}

If we use Planck's length as the unit of length, then in units of the
Boltzmann constant the entropy of a black hole of mass $M_{bh}$ is the
equivalent area formed by the product of its gravitational radius times the
size of the Universe. In this way we keep the entropy as an extensive
property. Also, the cosmological implication is here included through the
factor $ct$, an effect similar to the Mach's principle. This principle
relates the cosmological property of the mass of the Universe with the local
inertia properties of mass, i.e., cosmology with local objects. Relation (%
\ref{eq20}) has a cosmological parameter $ct$, the size of the Universe,
related to the local black hole. In this sense it is a Machean formula.

\section{\bf BLACK HOLE RADIATION.}

The characteristic wavelength of radiation from a black hole is about its
size, the gravitational radius $GM_{bh}/c^2$ . In the Bekenstein-Hawking
expression (\ref{eq17}) the horizon area, the square of the gravitational
radius, is the emitting source. In our case we propose the equivalent area $%
A_e$ in (\ref{eq20}) as the emitting source. In other words, we propose as
the characteristic wavelength of radiation $\lambda $ from a black hole the
size of the equivalent area:
\begin{equation}
\label{eq21}\lambda \approx \sqrt{A_e}\approx \sqrt{\frac{GM_{bh}}{c^2}\cdot
ct}
\end{equation}
which is much higher than the Hawking proposal. In particular, for the
primordial black holes one has
\begin{equation}
\label{eq22}\lambda \approx \sqrt{\frac{GM_{pbh}}{c^2}\cdot ct}\approx
10^8cm
\end{equation}
to be compared with $10^{-12}$ $cm$ deduced from the Hawking expression,
which is in the gamma ray range. Then, the wavelength being so large in our
case, the radiation of energy is much lower than in the Hawking case, about $%
10^{40}$ times lower. This means that black holes radiate much less than
thought and therefore they last longer in the Universe. If primordial black
holes were an important part of the dark matter they are still around us.
The fact that no gamma ray radiation is observed to account for them, a high
radiation predicted by the Hawking expression, is then no proof of their
absence. In our case the prediction is that if any, there should be a
radiation peak of a wavelength close to $10^8cm$. Interferometric methods
would be adequate to detect it, if there are enough primordial black holes
in the Universe. Then, for radiation from a black hole the antenna size is
given by (\ref{eq21}), a number much higher than its gravitational radius,
and linked to the Universe in a style similar to the Mach's principle.

\section{\bf CONCLUSION.}

A parallel gravitational uncertainty principle can be formulated using a
gravitational Planck's constant. Then, gravitational phenomena can be
treated as in the quantum mechanical case probably with very similar
formulation. The scale change is of the order of $10^{40}$ times smaller for
the gravity case today. The quantum of gravitational action has a mass about
$10^{-65}$ grams.\bigskip\

Entropy can be defined as an extensive property in terms of the equivalent
mass of the system using the gravity quanta. Hence, in any system a total
entropy concept for gravitational properties can be formulated by adding all
equivalent mass of its different energy components, in units of the quantum
of gravity. A corresponding black hole entropy can then be defined. Applied
to the primordial black holes one concludes that they radiate energy at a
much lower rate than in the Hawking model, $10^{40}$ times smaller. And
therefore they are still a candidate for the dark matter in the Universe.%
\bigskip\

The second law of thermodynamics for gravity, according to our view, is
summarized as follows: entropy increases linearly with time, the total
amount of entropy being the emitted equivalent number of gravity quanta
through the age of the Universe. Speculating about the future of this
Universe, taking physics beyond the theoretical, the time will stop when no
more gravity quanta are emitted, if the entropy arrives at a maximum and
remains constant.

\end{document}